\documentclass{PoS}
\usepackage{subcaption}

\title{Beyond the Standard Model Physics at the High Luminosity LHC}

\ShortTitle{BSM Physics at HL-LHC}

\author{\speaker{Sezen Sekmen} (for ATLAS and CMS Collaborations) \\ 
        Kyungpook National University \\
        E-mail: \email{ssekmen@cern.ch}}

\abstract{The High-Luminosity Large Hadron Collider (HL-LHC) is expected to deliver an integrated luminosity of up to 3000 fb$^{-1}$. The very high instantaneous luminosity will lead to about 200 proton-proton collisions per bunch crossing (pileup) superimposed to each event of interest, thus providing extremely challenging experimental conditions, which will be addressed by accompanying improvements in the decetors.  The sensitivity to find new physics Beyond the Standard Model (BSM) is significantly improved and will allow to extend the reach for SUSY, heavy exotic resonances, vector like quarks, dark matter and exotic long-lived signatures, to name a few.  This note summarizes several ATLAS and CMS studies performed to asses HL-LHC sensitivity to various BSM models and signatures.}

\FullConference{The 39th International Conference on High Energy Physics (ICHEP2018)\\
		4-11 July, 2018\\
		Seoul, Korea}

\begin{document}

The LHC will have collected about 300 fb$^{-1}$ of data until 2024, after which, it will go into an extensive upgrade phase.  
The accelerator will be improved to yield ten times more luminosity, which will lead to exciting physics prospects, but also to high radiation, large amounts of pile up and beam-induced backgrounds.  But detectors will also be upgraded to have higher granularity trackers and calorimeters; extended tracker and muon $\eta$ coverage; and improved track, vertex and muon triggering, timing, readout and radiation hardness, which will enable them to perform in these extreme conditions, and enhance their sensitivity.  Planning and construction for accelerator and detectors are already in progress.  Meanwhile, ATLAS and CMS also started to assess the rich physics potential for this exciting new era, called the high-luminosity LHC (HL-LHC).

ATLAS and CMS study a wide range of BSM models including supersymmetry, heavy resonances, vector like quarks, flavor anomalies, dark matter, dark sectors and long lived particles in view of the HL-LHC, either improving the existing searches to enhance sensitivity, or designing new searches to exploit new detector improvements or new analysis strategies.  Two methods are used to esablish sensitivity:
i) Run2 analyses are directly \emph{projected} to HL-LHC by scaling existing event counts to 14TeV cross sections and integrated luminosity, for different cases of systematic uncertainties and 
ii) \emph{Full analyses} based on existing or new methods are performed from scratch using dedicated samples produced with a parametrized simulation of the HL-LHC detector characteristics 
I present various example HL-LHC analysis in the following.

\begin{figure}[htbp]
\begin{center}
\begin{subfigure}{.49\textwidth}
\centering
\includegraphics[width=0.9\textwidth]{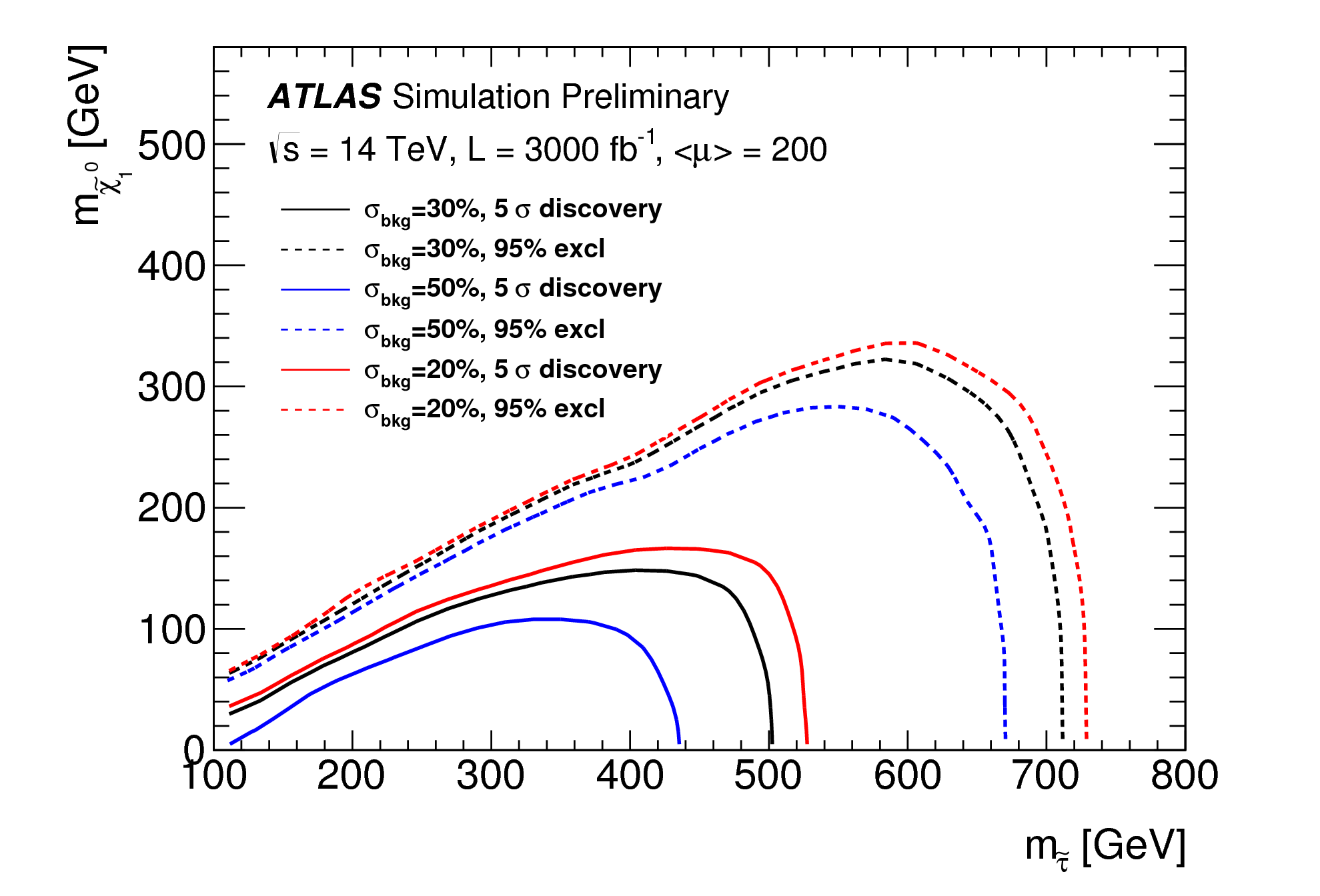}
\end{subfigure} 
\begin{subfigure}{.49\textwidth}
\centering
\includegraphics[width=\textwidth]{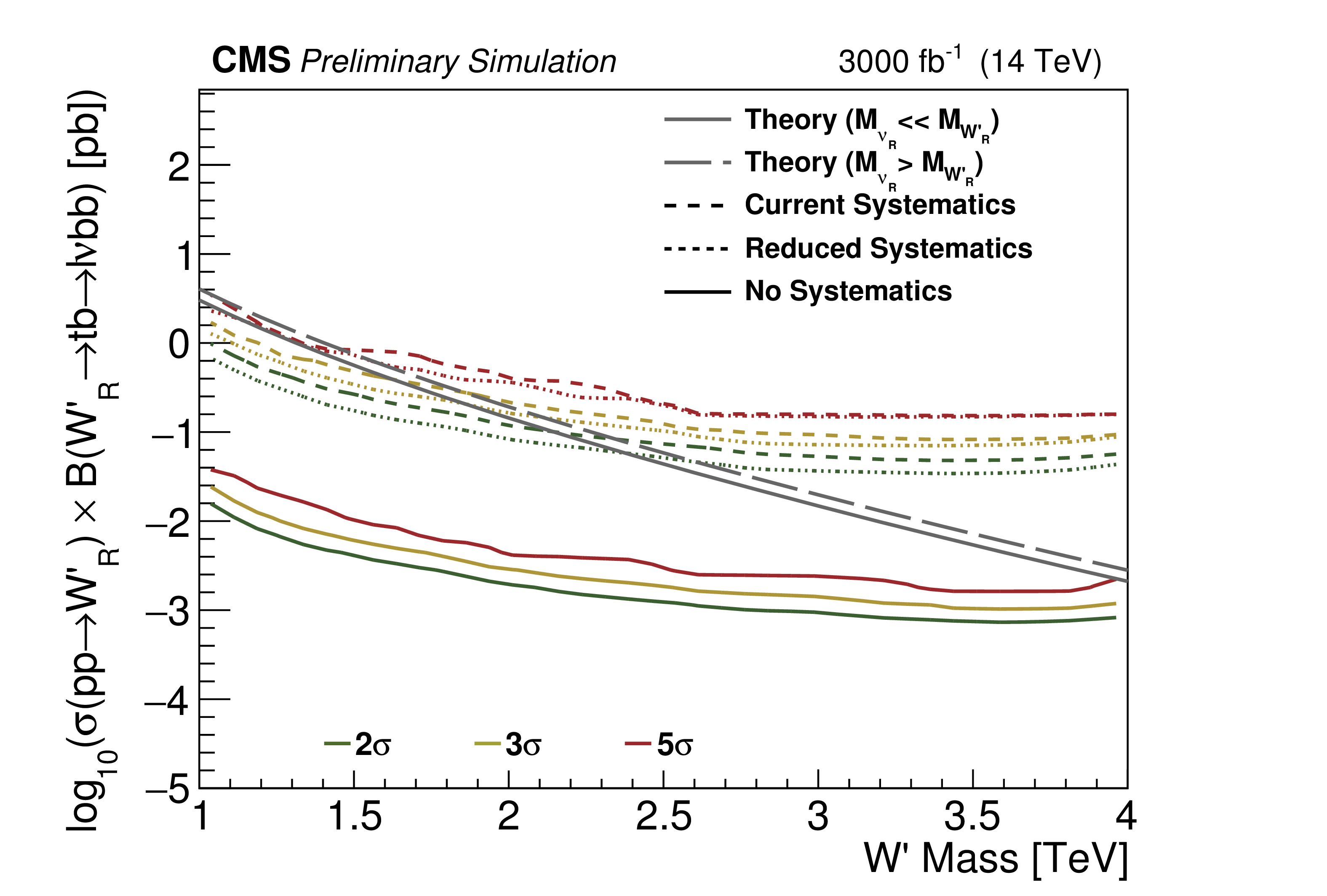}
\end{subfigure} 
\begin{subfigure}{.49\textwidth}
\centering
\includegraphics[width=0.9\textwidth]{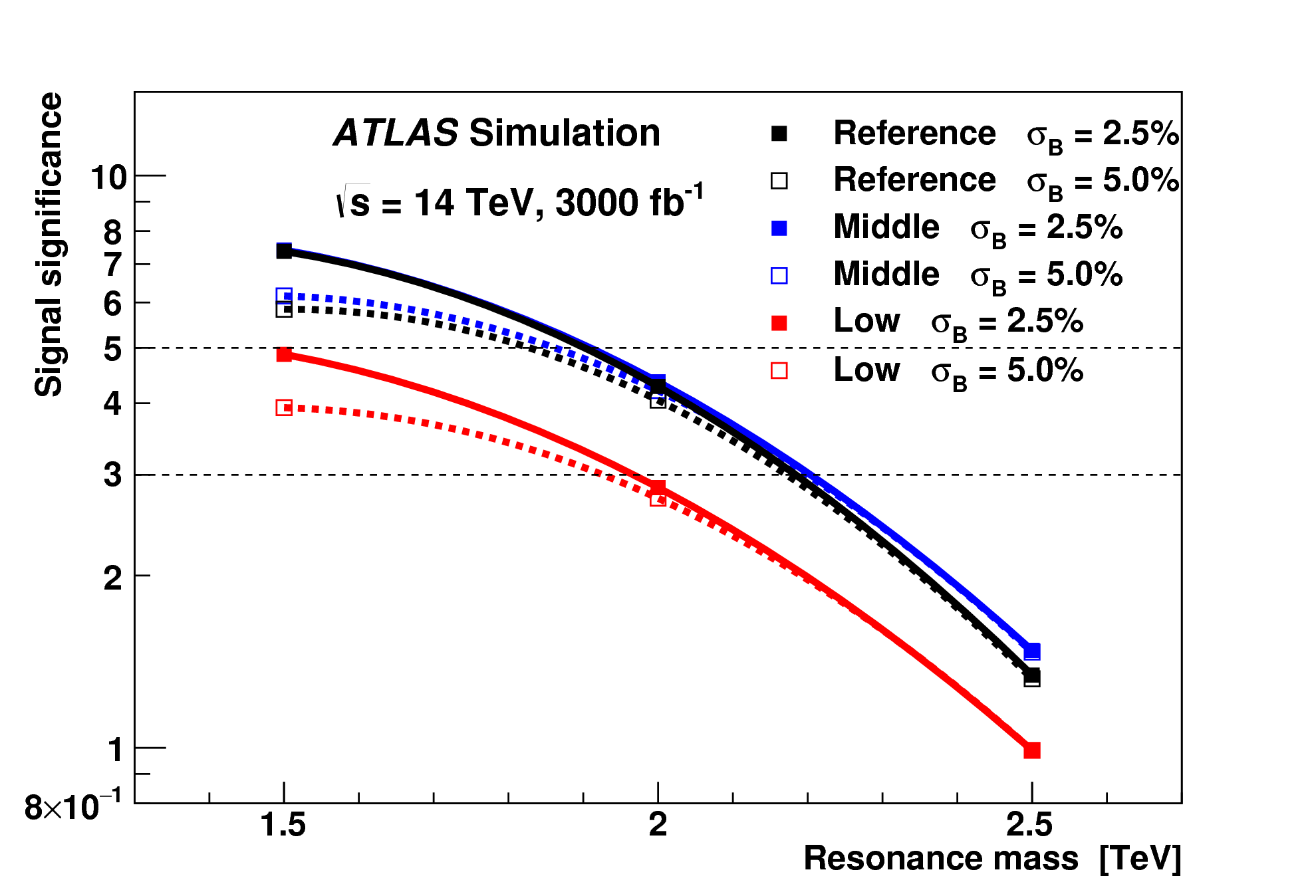}
\end{subfigure} 
\begin{subfigure}{.49\textwidth}
\centering
\includegraphics[width=0.8\textwidth]{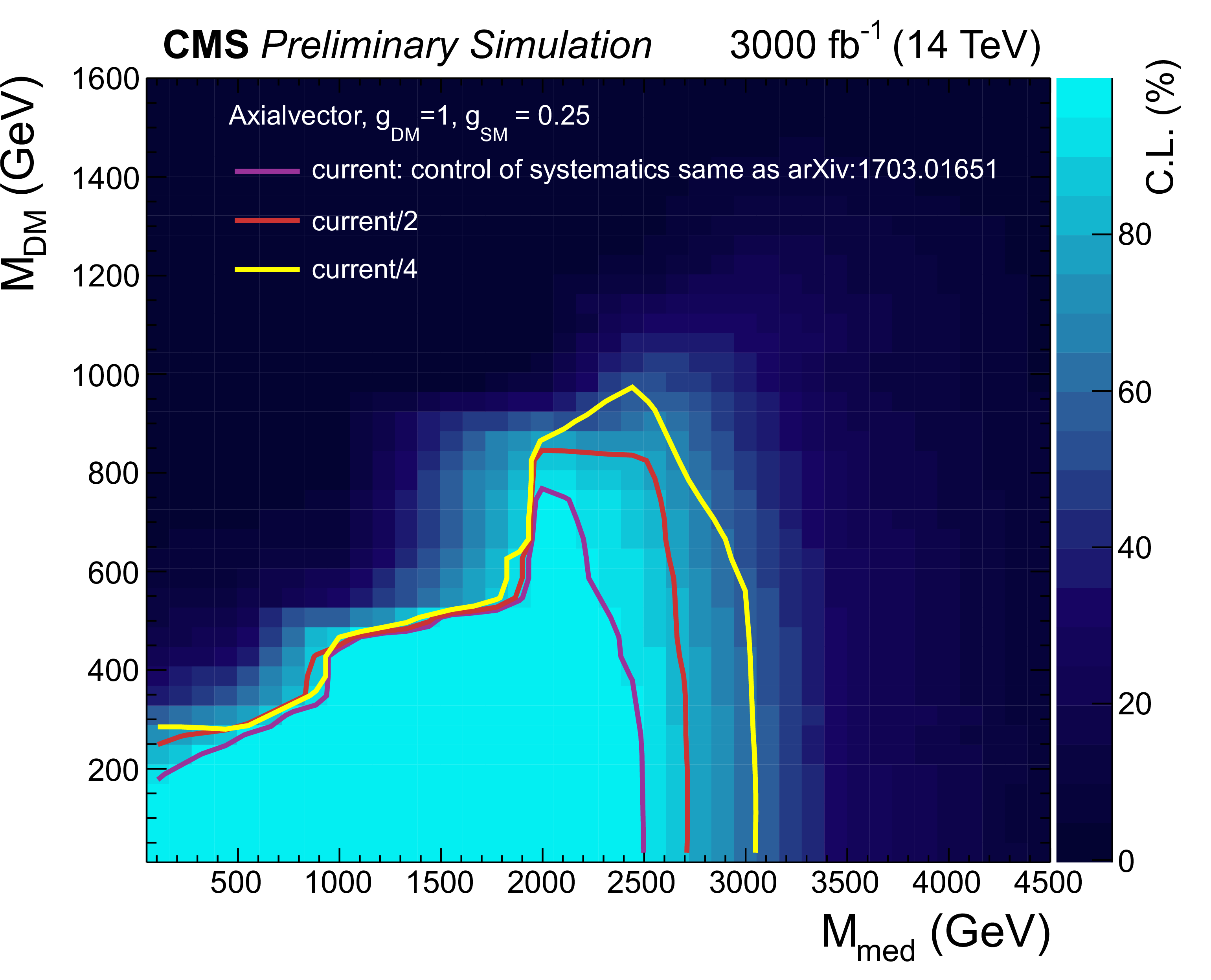}
\end{subfigure}
\begin{subfigure}{.49\textwidth}
\centering
\includegraphics[width=0.90\textwidth]{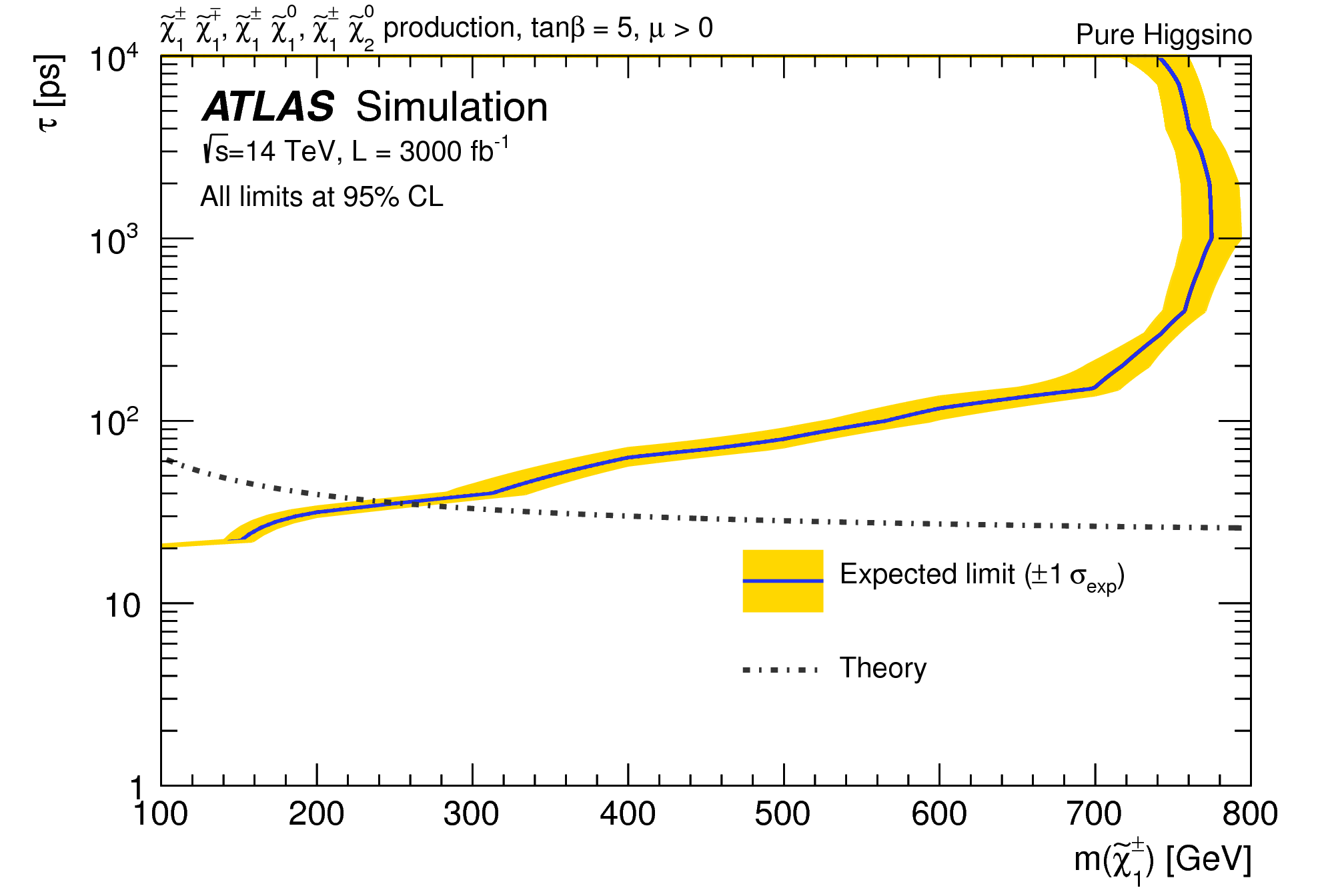}
\end{subfigure}
\begin{subfigure}{.49\textwidth}
\centering
\includegraphics[width=0.75\textwidth]{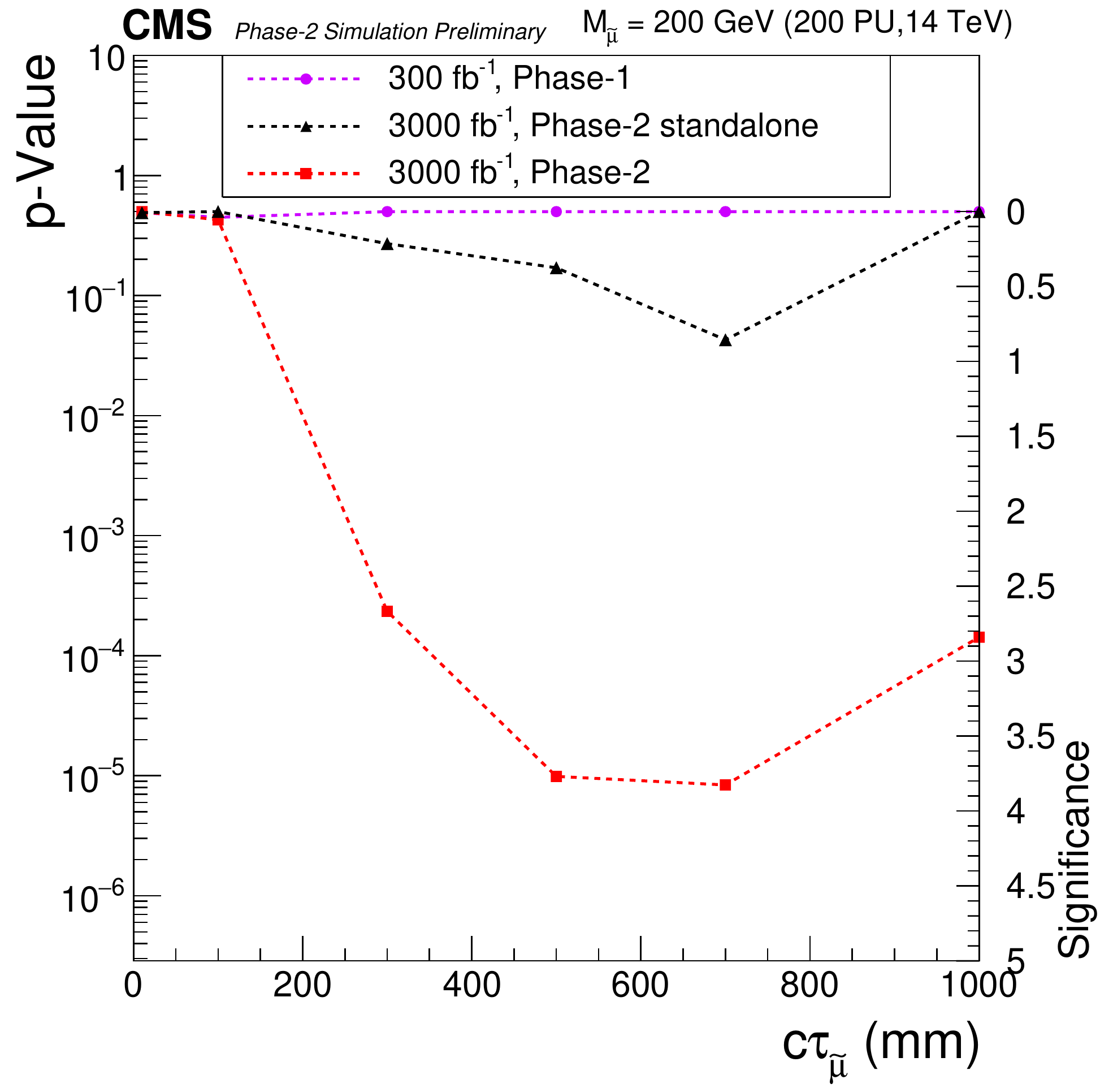}
\end{subfigure} 
\caption{Limits for example ATLAS and CMS HL-LHC BSM studies. See text for details. }
\label{fig:fig1}
\end{center}
\end{figure}

{\bf Supersymmetry (SUSY):}  Earlier work shows that gluinos and 1st/2nd generation squarks, bottom squarks and top squarks have up to 2.5~TeV, 2~TeV, 1.3~TeV and 1~TeV 5$\sigma$ HL-LHC discovery reach for the conventional searches.  Current HL-LhC studies focus on harder scenarios like heavy EW gaugino decays, compressed EW gauginos and sleptons, chargino pair production, light higgsinos, stau pairs, compressed stops and heavy stops/sbottoms.  One example is a direct search search for $\tilde{\tau}\tilde{\tau} \rightarrow \tau \tilde{\chi}_0^1 \tau \tilde{\chi}_0^1$ in the two opposite-sign hadronic taus plus missing transverse energy ($E_T^{miss}$) final state.  This scenario, having a dark matter relic density consistent with observation has low cross sections and acceptance.  A full analysis by ATLAS~\cite{atl-stau} shows that the current $\tilde{\tau}_L \tilde{\tau}_L, \tilde{\tau}_R \tilde{\tau}_R$ combined exclusion reach of 109~GeV can be extended to $\sim$700~GeV, and the $5\sigma$ discovery reach would be $\sim$500~GeV, as seen in Fig.~\ref{fig:fig1} (top left).  Another example is the new search for degenerate $\tilde{\chi}^\pm_2 \tilde{\chi}^0_4$ with mass $\sim M_2$ realized when the EWK parameters are ordered as $\mu < M_1 < M_2$.  The full CMS analysis in the 2 same sign leptonic W bosons $E_T^{miss}$ final state~\cite{cms-c2n4} predicts to exclude $m(\tilde{\chi}^0_4, \tilde{\chi}^0_4) \sim M_2 \simeq$ 900~GeV, leading to a potential to probe most of natural SUSY space.

{\bf Heavy resonances:}  These searches target heavy exotic resonances predicted by a wide range of BSM models.  One example is a full ATLAS $Z'$ search in the semileptonic $t\bar{t}$ channel, carried out by reconstructing boosted or resolved top quarks and taking their invariant mass~\cite{atl-zp}.  The mass reach for a $Z'$ in the TopColor model was estimated to be 4~TeV, greatly expanding the current exclusion of 2.1~TeV.  A similar $Z' \rightarrow t\bar{t}$ projection study by CMS in both hadronic and semileptonic channels~\cite{cms-bsm} extends sensitivity to 3.3~TeV for a narrow width $Z'$ resonance, and to 4~TeV for $Z'$ coming from Randall-Sundrum Kaluza-Klein (RS KK) graviton resonance.  A CMS projection study searches for $W'$ resonances predicted by the sequential SM decaying to $tb \rightarrow e/\mu + \nu bb$ in the $e/\mu +$ jets $+$ b jets $+E_T^{miss}$ channel by reconstructing the $tb$ invariant mass~\cite{cms-bsm}.  It extends the $3\sigma$ $W'$ exclusion limit from 2.7 to $>$4~TeV, and predicts a $5\sigma$ discovery up to 4~TeV as seen in Fig.~\ref{fig:fig1} (top right).  Meanwhile, a full ATLAS $W' \rightarrow \mu \nu$ study showed that the upgrade in the muon trigger sytem by adding inner barrel RPCs increased muon trigger acceptance from 70\% to 90\%~\cite{atl-wp}.  Another target scenario is that of di-Higgs resonances.  An ATLAS projection focused on KK excitation of gravitons in bulk RS models, decaying to two heavy neutral Higgs bosons, in turn, decaying to $b\bar{b}$~\cite{atl-hhbbbb}.  Higgsses were reconstructed via heavy Higgs tagging, using jet mass, substructure and subject $b-$tagging.  Searching for bumps in the di-boosted Higgs distribution extended the exclusion limit from 700~GeV to 2.5~TeV as seen in Fig.~\ref{fig:fig1} (middle left).  A projection by CMS studied the same resonance in the vector boson fusion production channel, using forward jet tagging, and obtained sensitivity up to 3~TeV~\cite{cms-hhbbbb}.

{\bf Vector-like quarks and leptons (VLQ, VLL):} A versatile HL-LHC VLQ and VLL search program is being developed to investigate production in heavy Higgs decays, measurement of chiral structure and other properties.  A full CMS search studied the EWK production of VLQ top partners as $gq \rightarrow Tbq' / Ttq$ with $T \rightarrow tH$ in the 1 lepton $+$ jets $+$ b jet(s) $+$ boosted H $+ E_T^{miss} +$ forward jet channel~\cite{cms-bsm}.  VLQ mass was reconstructed with a $\chi^2$ minimization.  While current searches have no sensitivity to this model, HL-LHC can reach above 1.5~TeV for certain cases.

{\bf Dark matter:} Many dedicated LHC DM searches are being refined for HL-LHC.  For these searches, simplified models of DM production are considered for spin-0 and spin-1 mediator cases, with DM candidate mass, mediator mass, mediator-DM coupling and mediator-SM coupling as free parameters.  Studies include DM $+$ monojet, DM $+$ heavy flavor (t, tt, bb, tttt) and DM $+$ mono Z, $\gamma$, and VBF production of DM.  As an example, Fig.~\ref{fig:fig1} (middle right) shows exclusion limits for an axialvector DM scenario by a CMS monojet $+ E_T^{miss}$ projection~\cite{cms-bsm}. 

{\bf Long lived particles (LLPs):} Detector upgrades at HL-LHC will improve trigger and object reconstruction and open a new era for the LLP searches.  One case is the search for a chargino $\tilde{\chi}^\pm_1$ almost degenerate with a neutralino $\tilde{\chi}^0_1$, decaying in the tracker to a very soft $\pi^\pm$ and a $\tilde{\chi}^0_1$, and leading to a disappearing track $+ E_T^{miss}$ signature.  An ATLAS full analysis~\cite{atl-pixtdr} showed that adding the new inner tracker strip detector (ITk), pushes the exclusion limits from 400 to 800 and 150 to 260~GeV for wino and pure higgsino (Fig.~\ref{fig:fig1} (bottom left)) $\tilde{\chi}^\pm_1$	.  Another full ATLAS analysis studied LLP decaying in ITk to stable particles forming displaced vertices and tracks~\cite{atl-pixtdr}.  It showed that larger volume and increased number of silicon layers in the ITk significantly improve displaced vertex acceptance for an R hadron decay~\cite{atl-pixtdr}.  A model-independent full CMS search studied displaced muons decaying outside the tracker~\cite{cms-mu}.  As triggering and reconstruction is done only in the muon system, additional hits in new muon layers and improved algorithms at HL-LHC will improve efficiency.  The analysis excludes GMSB smuons decaying as $\tilde{\mu} \rightarrow \tilde{G}\gamma$ up to 220~GeV and significantly improves the discovery reach as seen in Fig.~\ref{fig:fig1} (bottom right).  A different CMS full analysis studied displaced photons from $\tilde{\chi}^0_1 \rightarrow \tilde{G} \gamma$ employing the new MIP timing detector (MTD) ~\cite{cms-timing}.  Using the time of arrival of photons to MTD helped to discriminate signal and increased the sensitivity to short lifetimes and high masses.



\vspace{0.1cm}

{\bf Acknowledgements:} I thank my colleagues in the ATLAS and CMS collaborations for preparing a wide variety of HL-LHC BSM searches, the ICHEP2018 organizers, and the National Research Foundation of Korea for funding my participation through contact NRF-2008-00460.

\end{document}